\documentclass{article}

\usepackage{graphicx}       

\usepackage{physics}
\usepackage{lineno}
\usepackage{siunitx}
\usepackage[hidelinks]{hyperref}
\usepackage{cite}

\title{\textbf{A method to determine the M2 beam quality from the electric field in a single plane}}

\author{M.H. Griessmann, A.C. Martinez-Becerril, and J.S. Lundeen{*}  \\
  {\small Department of Physics and Nexus for Quantum Technologies, University of Ottawa,}\\
 {\small 25 Templeton Street, Ottawa, Ontario, K1N 6N5, Canada}  \\
  \texttt{ \small {*}jlundeen@uottawa.ca } \\
}
\date{}

\begin{document}
\maketitle

\begin{abstract}
Laser beam quality is a key parameter for both industry and science.
However, the most common measure, the $M^{2}$ parameter, requires
numerous intensity spatial-profiles for its determination. This is
particularly inconvenient for modelling the impact of photonic devices
on $M^{2}$, such as metalenses and thin-film stacks, since models
typically output a single electric field spatial-profile. Such a profile
is also commonly determined in experiments from e.g., Shack-Hartmann sensors, shear plates, or off-axis holography. We introduce and test
the validity and limitations of an explicit method to calculate $M^{2}$
from a single electric field spatial-profile of the beam in any chosen
transverse plane along the propagation direction. 
\end{abstract}

\section{Introduction}

Upon its invention in 1960, the laser immediately distinguished itself
from other light sources by its high quality spatial coherence, which
allowed it to propagate long distances in a pencil-like beam. Though
this beam quality is just one of the laser's many exemplary properties,
it has proven particularly useful to science and technology, for example
in distance measurements e.g., to the moon; high-resolution imaging,
cutting, machining, welding, and 3D printing; the characterization
of surface profiles such as the cornea; the construction of interferometric
sensors e.g., for gravitational waves; and long-distance power delivery
and communication. Nonetheless, it took another 30 years for the field
to settle on a standard performance metric for laser beam quality.
To this end, in 1990, Siegman drew attention to the $M^{2}$ parameter \cite{siegman1990,siegman1993defining,Siegman1998}, the ratio of the
product of the beam size in the near and far fields to the same product
for a diffraction limited beam \cite{siegman1993defining}. A reliable way to measure $M^{2}$
was codified in 1999 by the ISO standard 11146 \cite{ISO-1}, which
called for transverse spatial intensity-profiles to be recorded at
ten distances along the propagation axis. The $M^{2}$ parameter is
now routinely measured and reported in scientific research and included
in the advertised specifications of commercial lasers.

Since 1999, beam measurement tools and optical modelling have continued
to advance beyond intensity spatial-profiles. There are now established
methods and even commercial devices for experimentally measuring the
electric field spatial-profile of a beam. These include newer methods
such as off-axis holography \cite{digital_ImageFormation,Grilli:01} and direct measurement \cite{boyd2019quantum}, as well as time-tested devices such as Shack-Hartmann
sensors \cite{hartmann1904objektivuntersuchungen,shack1971production,schafer2002determination,sharma2014wave} and shear plates \cite{DigitalRecordingAndNumerical,schnars2014digital}. In addition,
with advances in computation, researchers and engineers, can now perform
accurate electric field modeling of a proposed device design using
finite-difference time-domain, transfer-matrix, or
finite element analysis methods. With such commercial or custom optical
simulation software one can predict the electric field profile produced
by a nano-antenna, or transmitted by a multilayer thin-film stack,
or reflected by a metalens, to give some examples. The $M^{2}$ parameter
is often the goal of such simulations, particularly of devices in
which new optical modes are excited e.g., large area optical fibers
\cite{wielandy2007implications,liao2009theoretical,Stutzki:14}, materials with a non-linear optical response \cite{Zhou:15}, and laser cavities \cite{hodgson1992beam,Jauregui:20}. In light of these advancements, intensity spatial-profiles may not always be the ideal choice for determining $M^{2}$.

Unlike an intensity profile, all the information about a laser beam's
quality is contained in its electric field spatial-profile in any
single transverse plane. This makes the numerous intensity profiles
and subsequent fitting of the ISO 11146 inconveniently circuitous, particularly
for optical simulations. Moreover, these numerous measurements in
the ISO standard are included, in a large part, to make the procedure
robust to measurement errors (e.g., background) and uncertainties,
which numerical modelling does not have. On top of this, for both
simulations and experiment, the ISO procedure is additionally problematic
since intensity profiles must be acquired at prescribed distances from
the beam waist. This requires a pre-characterization of the beam before
the numerous profiles are obtained, i.e., to find the approximate
location of the beam waist and the Rayleigh range. The complexity
of the ISO standard has motivated work over the last decades to simplify
the experimental procedure to obtain $M^{2}$; for an overview see
\cite{marshall2012handbook,forbes2014laser}. 

Early papers on $M^2$ \cite{lavi1988generalized,weber1992some, du1992coherence,siegman1991defining,gao2005characterization} focused on characterizing $M^2$ through intensity measurements. Some papers  \cite{liao2009theoretical, weber1992some, du1992coherence, Yoda:06} have developed theory based on the electric field distribution to derive formulae for $M^2$ for specific beam shapes with analytic forms. Recently, other works have developed experimental methods to obtain $M^2$ from the electric field, albeit indirectly. For example, \cite{Flamm:12,du2013real,du2016complex} numerically propagated the electric field to different planes to replicate the ISO standard method, and \cite{schafer2002determination} demonstrated a method to determine $M^2$ using a Shack-Hartmann sensor along with near and far-field intensity profiles. Others measure the modal decomposition of a beam in order to calculate $M^2$, e.g. \cite{Schmidt:11}. In this work, using a position-angle phase-space picture, we motivate the fundamental meaning of $M^2$ and use that to introduce a method to find $M^2$ that only requires the electric field at one plane. This eliminates the need for numerical propagation or numerical projection onto a modal basis. We numerically validate such a method for two nontrivial test beams.

The rest of this paper is organized as follows. In Section \ref{sec:two},
we start by defining $M^{2}$ and outline the method in  ISO 11146
\cite{ISO-1} (which we call the "ISO standard" from hereon). We then introduce an angle-position phase-space
for a beam's state and relate $M^{2}$ to the state's area, which we
show is conserved under paraxial propagation (i.e., the range of angles $<< \SI{1}{rad}$). We give explicit formulas
to calculate $M^{2}$, the beam waist $w_{0}$ and its location $z_0$, the
beam's angular width  $\Delta\theta$, and the Rayleigh range
$z_{R}$ from the electric field profile at one plane. We call this the \textit{ covariance method}. In Section
\ref{sec:three}, we validate our covariance method by numerically comparing
its predictions for $M^{2}$ to the ISO method for two non-trivial
beam shapes.  In Section \ref{sec:four}, we discuss the advantages
and drawbacks of our covariance method and the prospects for generalizing it
to more complicated beams. While many of the concepts in this paper
have been introduced elsewhere, they have not been connected in a simple
way to allow for easy use. For those solely interested in applying
our covariance method, in Section \ref{subsec:Q_matrix} we summarize our results
and provide a detailed straightforward recipe for determining $M^{2}$
from a single electric field profile of the beam anywhere.

\section{Theoretical description of laser beams\label{sec:two}}

\subsection{Definitions and conventions}
For simplicity we will consider beam profiles in one transverse dimension
only, although results generalize to two dimensions. The $z$ axis is the propagation direction while $x$-axis
is the transverse direction, along the beam's spatial profile.  As a beam propagates along $z$, generally its width in position $w\left(z\right)$
will change while its angular width $\Delta\theta$ will be constant. We use the standard width conventions $w\left(z\right)=\Delta x\left(z\right)=2~\sigma_{x} \left(z\right)$ and $\Delta\theta=2~\sigma_{\theta}$, where the $\sigma$ are the standard deviation of corresponding intensity distribution (i.e., in $x$ or $\theta$). We take
$z=0$ as the position of the beam waist  $w_{0}$, the beam's minimum
$w\left(z\right)$ over all $z$.  It is important
to note that, except for Gaussians, the $w_{0}$ width here differs
from the definition of the beam waist width common in Gaussian optics, the $1/e^{2}$ intensity
half-width. The Rayleigh range $z_{R}$ is defined as the distance
from the waist (taken as $z=0$) at which the beam has increased in
width to $w\left(z_{R}\right)=\sqrt{2}w_{0}$. 

\subsection{Definition of the $M^{2}$ parameter}

The beam or mode quality $M^{2}$, also known as the
beam propagation factor, is defined in terms of the beam waist $w_{0}$ and angular width $\Delta\theta$.  The product of these two widths divided
by the same product for a Gaussian beam is the definition of $M^{2}$.
Since for any Gaussian this product equals $\lambda/\pi$ \cite{Siegman1998}
we have,
\begin{equation}
M^{2}=\frac{\pi}{\lambda}w_{0}\Delta\theta.\label{eq:M2basic}
\end{equation}
Whereas the product $w\left(z\right)\Delta\theta$
will change as the beam propagates, since $w_{0}$ and $\Delta\theta$
are independent of $z$ their product will not change. They are characteristics of the beam as a
whole, as is the beam quality $M^{2}$.

\begin{figure}[hbt!]
  \begin{center}
 \hspace{5mm}  (a) \hspace{52mm} (b) \\
  \includegraphics[width=6cm]{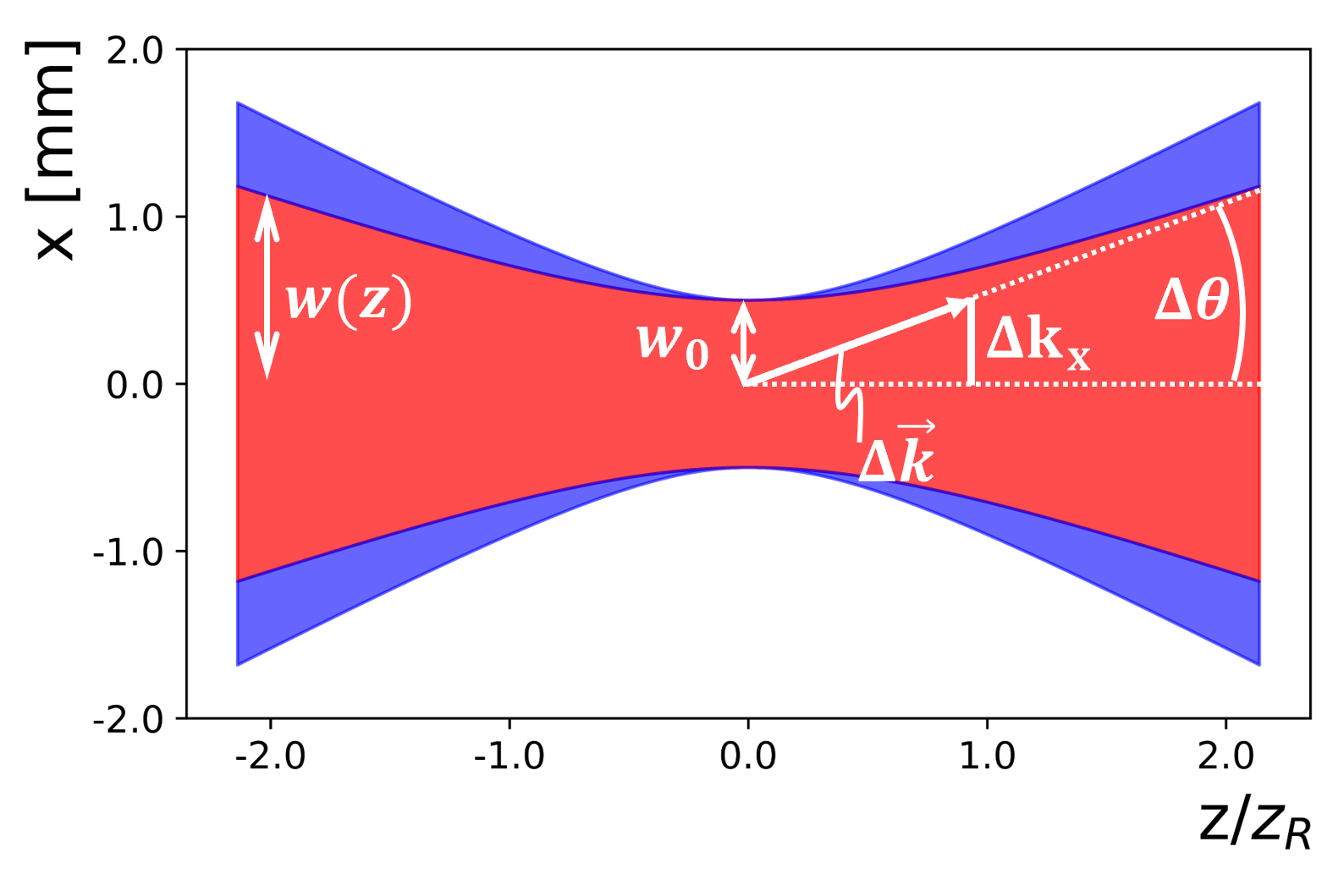}
  \includegraphics[width=6cm]{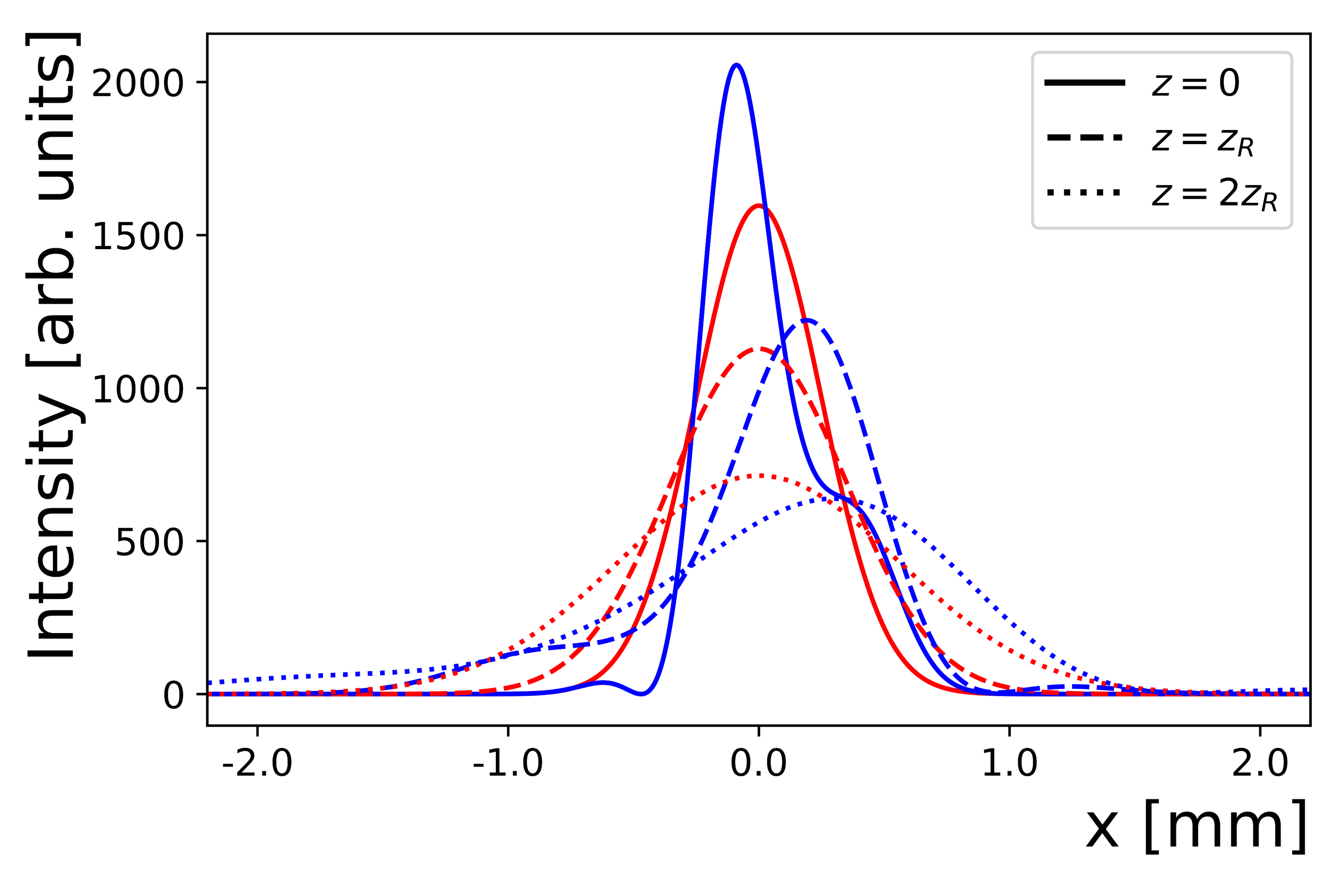} \\
 \hspace{5mm}  (c) \hspace{52mm} (d)\\
  \includegraphics[width=6.cm]{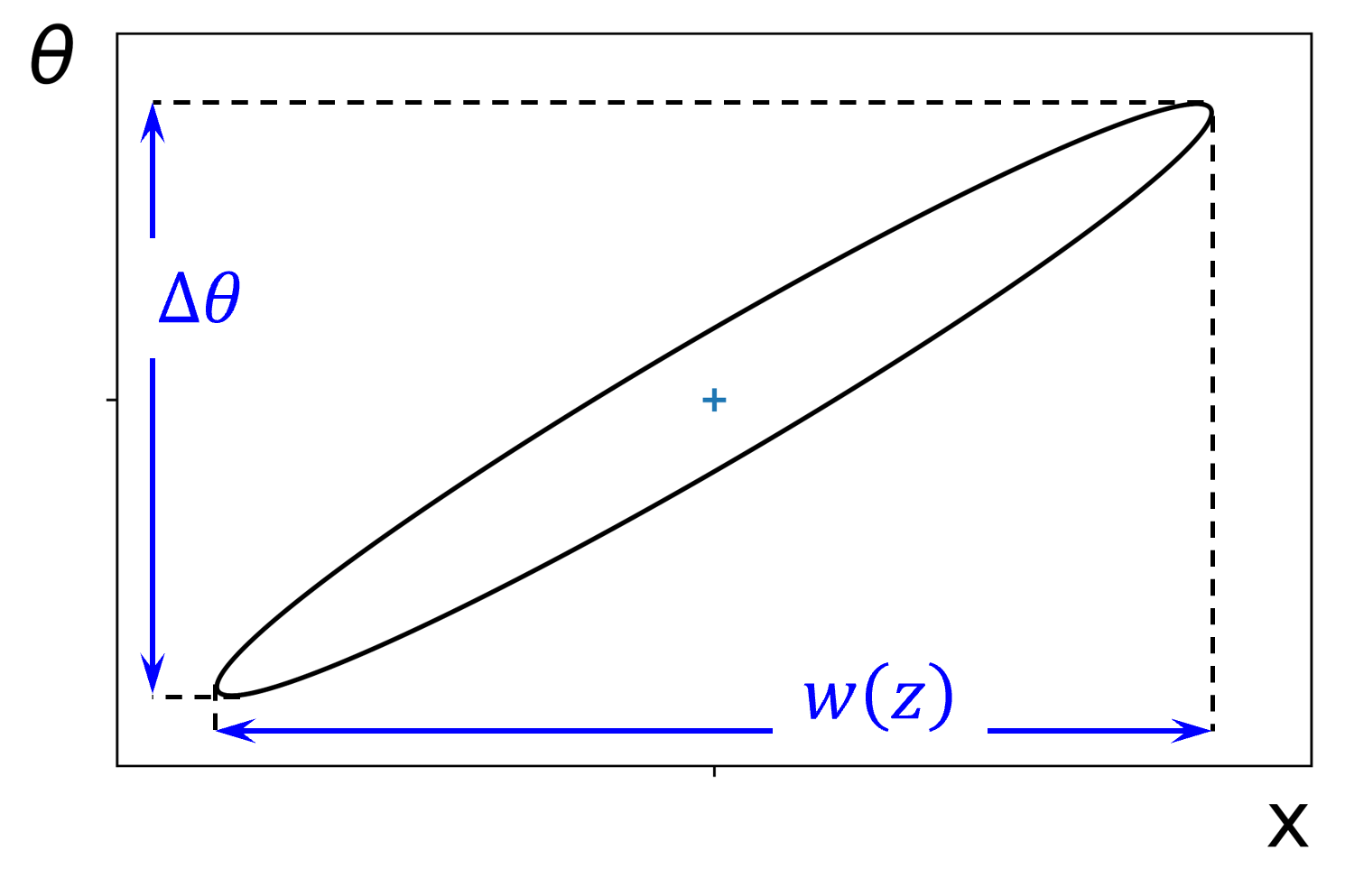}
  \includegraphics[width=6.cm]{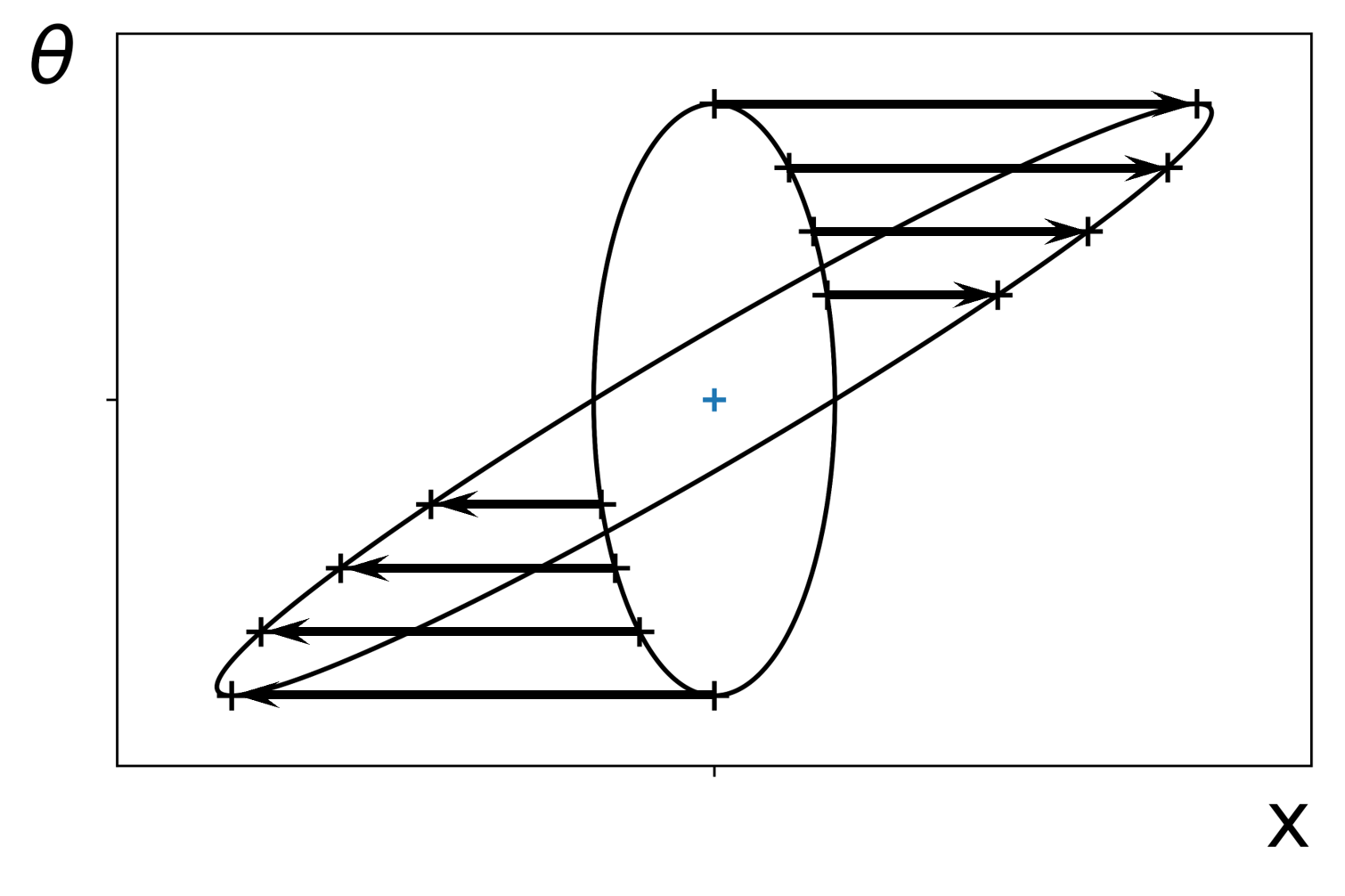}
  \caption{The evolution of paraxial beams under propagation. (a) Sketch showing the beam width along $z$ for a Gaussian beam (red, $M^2=1$) and combination of first four Hermite-Gauss modes (blue, $M^2=1.5$) with $w_0=$ \SI{500}{\um}, $\lambda=$ \SI{400}{\nm}. The $z$-values are in units of the Gaussian's Rayleigh distance $z_R = $ \SI{1.96}{\m}. (b) Normalized intensity profiles of the same beams as in (a) at three $z$-positions. (c) The phase-space ellipse representing the beam state at a general $z$-position. (d) The $x$-shear resulting from paraxial propagation of the beam. Blue cross indicates origin.
    \label{fig:beam_basics}
  }
  \end{center}
\end{figure}

As we will show, the value of $M^{2}$ is bounded by the Heisenberg uncertainty principle. 
As can be seen in Fig. \ref{fig:beam_basics}a, the relation between
the widths in angle ($\Delta\theta$), transverse wavevector
($\Delta k_{x}$) and transverse momentum ($\Delta p_{x}$) of a photon
can be expressed as 

\begin{equation}
\Delta p_{x}=\hbar\Delta k_{x}=\hbar k\sin{\Delta\theta}\approx\hbar k\Delta\theta,\label{eq:Relation_angle_momentum}
\end{equation}
where $k=2\pi$/$\lambda$ is the total wavevector magnitude, $\lambda$
is the wavelength, and the small angle approximation was applied. Using Eq.~(\ref{eq:Relation_angle_momentum}), one can rewrite $M^2$ as
\begin{equation}
M^{2}=\frac{1}{ 2 \hbar}\Delta x\Delta p_{x}\geq1,\label{eq:M2_inequality}
\end{equation} 
where the inequality comes from the Heisenberg uncertainty
principle: $\sigma_{x}\sigma_{p_x}\geq\hbar/2$ or, equivalently, $\Delta x\Delta p_{x}\geq2\hbar$. Gaussian beams are minimum uncertainty states and thus saturate 
 inequality (\ref{eq:M2_inequality}), $M^{2}=1$, while all other beams give strict inequality.

Although Eq.~(\ref{eq:M2basic}) for $M^{2}$ seems quite simple, determining its value is deceptively difficult. The angular width $\Delta\theta$ can be found from the beam
width far from the waist, i.e., in the far-field.
In contrast, to measure the beam waist $w_{0}$ requires that one knows its $z$-location. This is typically not known a priori, and it is seldomly known
with good accuracy. The ISO standard avoids this accuracy issue
by recommending measuring the beam width $w\left(z\right)$ in five different
$z$-planes on each side of the beam waist, half of which should lie
within one Rayleigh range $z_{R}$ and the rest at least two Rayleigh
ranges away from it. While one is free to choose the precise locations
of the planes, to do so one still must approximately know $z_{R}$
and the location of the beam waist. A rough pre-characterization can
take the place of this a priori information, but this adds complexity
to the procedure. The basic definitions (Eq.~(\ref{eq:M2basic})) are hardly
applicable in either experiment or simulation.

\subsection{The connection between phase-space and the electric field}

We now introduce a phase-space picture of the beam in order to justify
a simpler method to determine $M^{2}$. Appropriately, the 2D phase-space is made up of position $x$ on one axis and angle $\theta$
on the perpendicular axis (though the latter could equally well be
$k_{x}$ or $p_{x}$, as explained in the last section). In this phase-space, the beam profile at any $z$-position can be visualized as
a tilted ellipse (see Fig. \ref{fig:beam_basics}c).  The length of the
projection of the ellipse on the $x$-axis is the full-width intensity
standard deviation, $w\left(z\right)=\Delta x$. Similarly, the projection on
the $\theta$-axis is the angular beam width, $\Delta\theta$ .

When the
beam has reached its waist (taken as $z = 0$), the ellipse is aligned
with the phase-space axes and its full-width in $x$ is minimized,
equalling $w_{0}$. The area $\mathcal{A}$ of the ellipse is proportional
to its width times its height, $\mathcal{A}=\tfrac{1}{4}\pi w_{0}\Delta\theta$.
In turn, using Eq.~(\ref{eq:M2basic}), we find that the beam quality
is proportional to the ellipse area, 
\begin{equation}
M^{2}=\frac{4\mathcal{A}}{\lambda}.\label{eq:Area}
\end{equation}

As the beam propagates away from the waist location, correlations between position and angle form
and the ellipse becomes tilted. This follows from simple geometry;
waves traveling at at angle $\theta$ will increase in position as
$x=\theta z$ in the small angle approximation, creating a correlation
between $x$ and $\theta$. This shears the ellipse along the $x$-direction
in phase-space \cite{Alonso2011} as seen in Fig.~\ref{fig:beam_basics}d.
(Conversely, momentum conservation in free space ensures that each
constituent wave's angle $\theta$ is constant.) However, since the ellipse is now tilted, these
widths are distinct from the width and height of the ellipse itself
i.e., along its semi-minor and semi-major axes. Crucially, in geometry,
a shear transformation always preserves area \cite{crampin1986applicable,hazewinkel2012encyclopaedia} and, thus, $M^{2}$. This implies that Eq.~(\ref{eq:Area}) is true for all $z$, not just at the waist \cite{dodonov1989universal,simon2000optical,dodonov2000universal}. Thus,  the problem of determining $M^{2}$ from information in a
single $z$-plane reduces to finding the area of the corresponding tilted
ellipse. 

\subsection{The covariance matrix}\label{subsec:matrix_method}
We gather the parameters of a tilted ellipse centered at the origin 
(i.e., the ellipse equation, $ax^{2}-2bx\theta+c\theta^{2}=ac-b^{2}$)
into a symmetric matrix \cite{pettofrezzo1966matrices}, a method from geometry known as the \textit{matrix of quadratic form.} This matrix sets the ellipse aspect ratio, orientation
of its axis, and its size. In the context of beam widths, the matrix
of quadratic form is the following covariance matrix,
\begin{equation}
Q\left(z\right)=\begin{bmatrix}\expval{ x^{2}} & \frac{1}{2}\expval{ x\theta+\theta x}\\
\frac{1}{2}\expval{ x\theta+\theta x} & \expval{\theta^{2}}
\end{bmatrix}\equiv\begin{bmatrix}a & b\\
b & c
\end{bmatrix},\label{eq:covariance matrix}
\end{equation}
where  $\expval{}$ is analogous to an average or expectation value but evaluated using the complex distribution $E\left(x;z\right)$, the transverse profile of the beam's electric field at a plane $z$. We give further detail on calculating  $\expval{}$ at the end of this subsection and explicit expressions for $Q$ will be given in Section \ref{subsec:Q_matrix}. For now, we point out that the diagonals of $Q$ are the variances,
$\expval{ x^{2} }=\sigma_{x}^{2}=w^{2}/4$ and $\expval{ \theta^{2}}=\sigma_{\theta}^{2}=\left(\Delta\theta\right)^{2}/4$
for a beam with $\expval{ x } = \expval{ \theta } = 0$. When the
beam has reached its waist $z=0$, the ellipse is aligned with the
phase-space axes and $Q$ is diagonal. As explained in the previous subsection, at other $z$, correlations exist between angle and position. These are the off-diagonal covariance terms in $Q$. The $Q$ matrix is closely related to beam matrices defined in terms of the Wigner
function; namely, it is the Weyl transform of the matrices in \cite{Bastiaans}
and Part 3 of the ISO standard \cite{ISO-3}. Unlike those matrices, $Q$
is directly computed  from the electric field distribution $E\left(x;z\right)$
in a single $z$-plane. 

The advantage of this matrix formulation is that the determinant of the matrix $Q$ is proportional to  the ellipse area, regardless of any tilt. The ellipse area is now simple to find by $\mathcal{A}=\pi\sqrt{\det Q}$. Combining this with Eq.~(\ref{eq:Area}) we arrive at our central idea,
the determinant of the covariance matrix is directly related to $M^{2}$
via 
\begin{equation}
M^{2}=\frac{4\pi}{\lambda}\sqrt{\det Q}=\frac{4\pi}{\lambda}\sqrt{ac-b^{2}}.\label{eq:M2fromdet}
\end{equation}
This relation is valid at arbitrary $z$-planes as long as the (paraxial)
beam propagates only through first-order optical systems (e.g., spherical mirrors and lenses) or freely
in space \cite{Bastiaans}.

In summary, one can evaluate $M^{2}$ using
the elements of the covariance matrix Eq.~(\ref{eq:covariance matrix}). These expectation values use the complex
field $E\left(x;z\right)$ (in analogous way to the quantum wavefunction). That
is, the expectation value of a general operator $v$ acting on $E$
is defined by $\expval{ v }=\tfrac{1}{n}\int{E^{*}{v}Edx}$, where
$E^{*}$ is the complex conjugate of the electric field, and $n=\int{|E\left(x;z\right)|^{2}dx}$
is a normalization factor. The angle operator is $\theta=-\frac{i}{k}\frac{\partial}{\partial x}$,
while $x$ is simply a multiplication by $x$. Note, the action
of these two operators changes if their order changes, which may motivate
why each off-diagonal in the covariance matrix incorporates both orderings.
In general, the phase-space ellipse might not be centred at the origin,
i.e., $\expval{ x } \neq0$ and/or $\expval{\theta}\neq0$.
We can account for this by substituting $x$ and $\theta$ in the covariance
matrix (Eq.~(\ref{eq:covariance matrix})) with $x- \expval{x} $ and $\theta-\expval{\theta}$, respectively. We give explicit formulae for elements $a,b,$ and $c$ for such offset beams in Appendix \hyperref[app:Dnon_paraxial]{D}.
For example, the beam width can be evaluated as 
\begin{equation}
\Bigl(w\left(z\right)\Bigr)^{2}=\frac{4}{n}\int{|E\left(x;z\right)|^{2}\Bigl(x-\expval{x}\Bigr)^{2}dx}. \label{eq:beam_width}
\end{equation}

In Appendix \hyperref[app:Aparaxial]{A}, we analytically model the propagation of the $Q$ matrix. We bring these definitions together to provide an explicit formula for $M^{2}$ and other key beam parameters in  Section \ref{subsec:Q_matrix}. 

\subsection{Recipe to compute $M^{2}$ and other beam parameters from the electric
field \label{subsec:Q_matrix}}
The goal of this section is to be self-contained and provide a straightforward set of steps to calculate $M^{2}$ and other key beam parameters directly from the transverse profile of the beam's scalar electric field $E\equiv E\left(x;z\right)$ that one has obtained in any plane $z$ along the propagation direction. To do so, we use Eq.~(\ref{eq:M2fromdet}) and the covariance matrix
Eq.~(\ref{eq:covariance matrix}) , which was found under the assumption that $\expval{ x } = \expval{ \theta } = 0$. This assumption is valid for most optical simulations since the incoming beam is ideal and usually travelling centered on the $x$-axis and because the optical device usually is symmetric about $x=0$. We give explicit formulae for elements $a,b,$ and $c$ for the more general case, offset beams, in Appendix \hyperref[app:Dnon_paraxial]{D}.
Reviewing, the beam must be paraxial
with angles to $z$-axis much less than 1 rad; as usual, the width
convention for the beam waist $w_{0}$ and angular spread $\Delta\theta$
is twice the intensity standard-deviation (e.g., $w_{0}=2\sigma_{x})$;
and the beam quality is 

\begin{equation}
    M^{2}\equiv\frac{\pi}{\lambda}w_{0}\Delta\theta=\frac{4\pi}{\lambda}\sqrt{ac-b^{2}}.\label{eq:M2_review}
\end{equation} 

To evaluate this formula, one uses the covariance matrix elements,
all of which are real-valued: 
\begin{eqnarray} \label{eq:abc}
a & = & \frac{1}{n}\int{x^{2}|E|^{2}dx},\\
b & = & -\frac{i\lambda}{4\pi}\left(1+\frac{2}{n}\int{xE^{*}\pdv{E}{x}dx}\right),\nonumber \\
c & = & -\left(\frac{\lambda}{2\pi}\right)^{2}\frac{1}{n}\int{E^{*}\pdv[2]{E}{x}dx},\nonumber \\
n & = & \int{|E|^{2}dx},\nonumber 
\end{eqnarray}
where we have used $\partial/\partial x \left(xE\right)=x\partial E/\partial x+E$  to simplify $b$.
One can save some computational effort my noticing that the second and first integrals in $a$ and
$c$, respectively, also appear in $b$. 

From the $a,b,$ and $c$ matrix elements, given above we can also
find other key beam parameters, such as the beam's angular width
$\Delta\theta$, Rayleigh range $z_{R}$, beam waist $w_{0}$,
and waist's location relative to the plane of the measured field,
$\Delta z$. In Eqs.~(\ref{eq:Angle_width_c}-\ref{eq:rayleigh_range_c}) in Appendix \hyperref[app:Aparaxial]{A}, we
show that
\begin{equation}
\Delta z=\frac{b}{c},~~~~~~w_{0}=2\sqrt{a-b^{2}/c},~~~~~~\Delta\theta=2\sqrt{c},~~~~~~z_{R}=\frac{\lambda}{4\pi c}.\label{eq:parameters}
\end{equation}
 Thus, like the beam quality, these parameters can be determined from
the electric field profile at one plane.

The formulas in this section are the main results of this paper. The
next section numerically verifies that they are correct, and also
provides further details for the numerical evaluation of the electric
field integrals in the $a,b,$ and $c$ parameters.

\section{Simulation and results\label{sec:three}}

In this section, we compare the $M^{2}$ found from our covariance method to the 
ISO standard via a numerical simulation of propagation of two beams with
nontrivial electric field profiles. We calculate $E\left(x\right)$ at an initial plane ($z=0$) using the analytic
form of the test beam $E\left(x\right)$ and then numerically propagate it to
find $E\left(x;z\right)$ and the corresponding intensity profile $I\left(x;z\right)=\left|E\left(x;z\right)\right|^{2}$
at the ten requisite planes described below. This propagation is detailed in Appendix \hyperref[app:BFT_prop]{B} and contains no paraxial approximation.

For the implementation of the
ISO protocol, we calculate the standard deviation of the transverse intensity profile $I\left(x;z\right)$ to find the beam width $w\left(z\right)$ using Eq.~(\ref{eq:beam_width}). We do so at five planes within half a Rayleigh distance on either
side of the beam waist and at five planes in the range of four to
five Rayleigh distances. These widths are fit to Eq.~(\ref{eq:ISO}) to
find $M^{2}$. 

At each of these planes,  we calculate
the entries, $a$, $b$, and $c$ of the covariance matrix $Q$ from $E\left(x;z\right)$
by Eq.~(\ref{eq:abc})  and use them to calculate $M^2$  using Eq.~(\ref{eq:M2fromdet}).
We compare these ten values of $M^{2}$ from our covariance method, which should
all be identical, to the single value from the  ISO protocol. In Appendix \hyperref[app:CNumericalDetails]{C} we give some details of the numerical
implementation.

\subsection{Numerical comparison of the ISO and covariance methods} \label{subsec:comparison}

For the following test beams, $\lambda = $ \SI{400}{\nm}, we use position range of length $L=$ \SI{200}{\mm} and a position grid density $\delta x=\lambda/2$. The ten planes range from $z=-7$ to \SI{7}{\m}.

\begin{figure}[hbt!]
  \begin{center}
\hspace{5mm}   (a) \hspace{52mm} (b) \\
  \includegraphics[width=6cm]{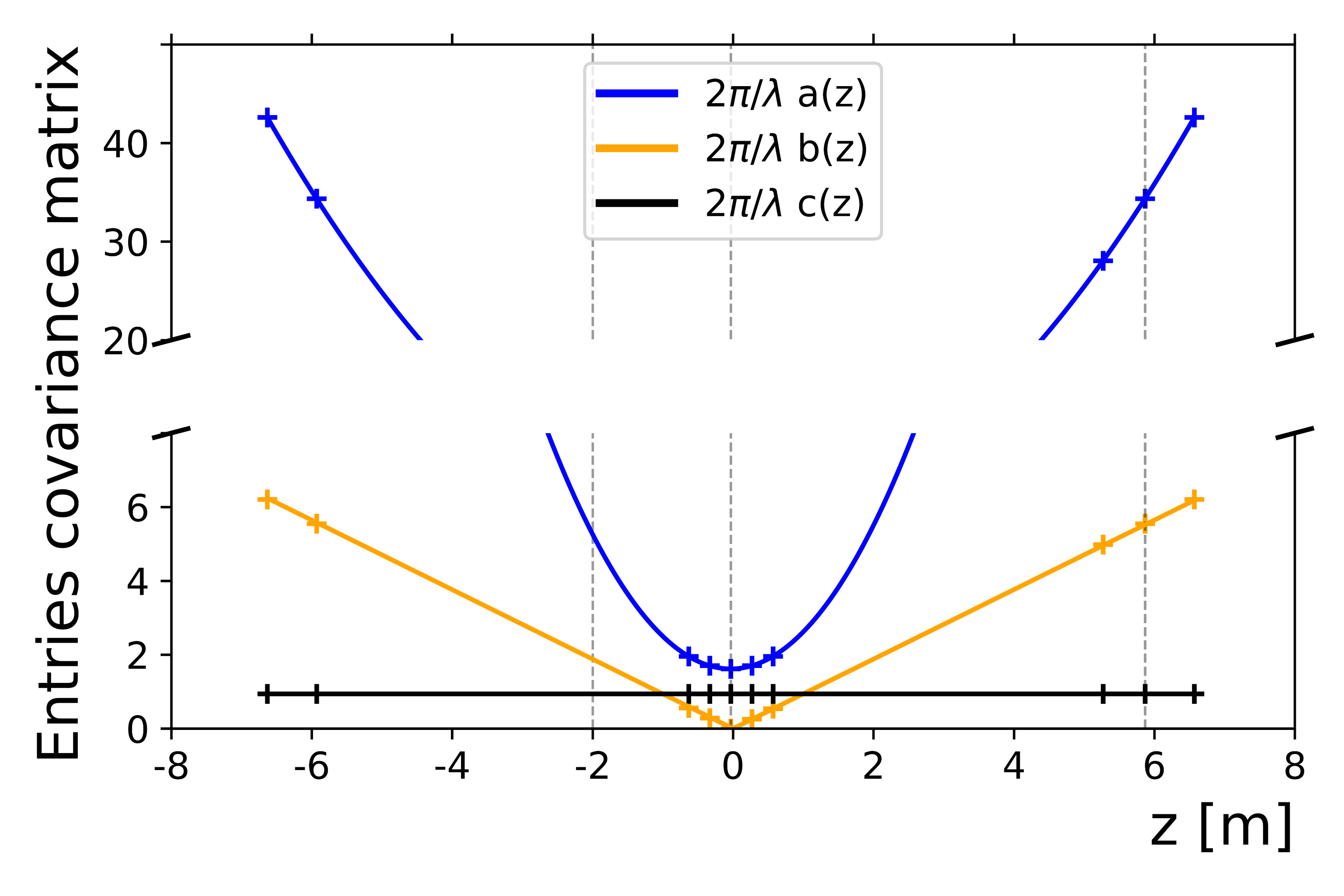}
  \includegraphics[width=6cm]{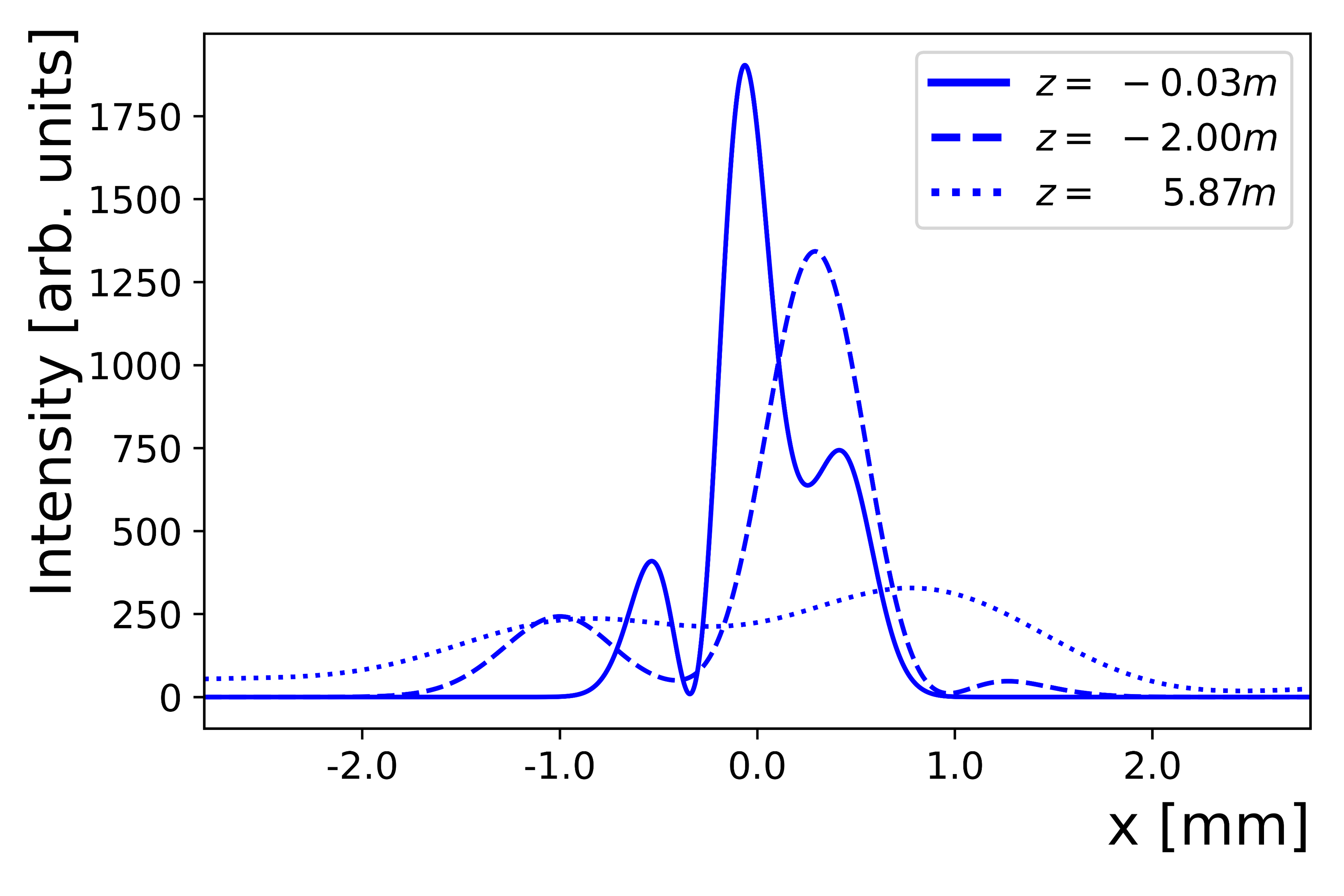}\\
 \hspace{3mm}  (c)\\
  \includegraphics[width=6cm]{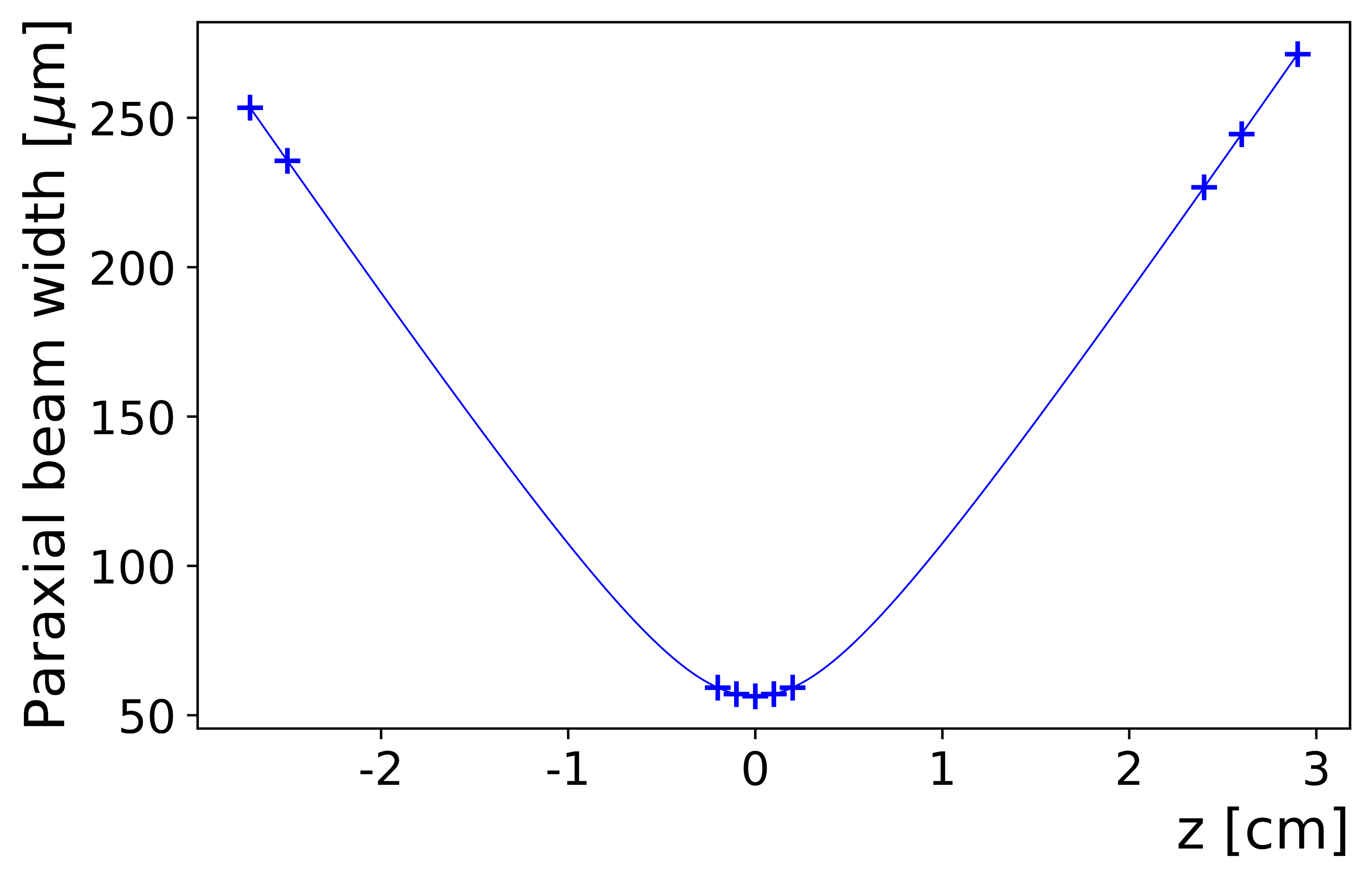}
  \caption{Analytic and numerical propagation of test beams. For a beam that is the superposition of the first four Hermite-Gauss  modes given in the main text ($w_0=$ \SI{642}{\um}, $\lambda=$ \SI{400}{\nm} $M^2=2.466$ given by Eq.~(\ref{eq:M2HG}) ): (a) points are the scaled entries of the covariance matrix for the beam numerically propagated (by Eq.~(\ref{eq:angular_spectrum})) to each $z$ distance and curves are the analytically propagated elements (by Eq.~(\ref{eq:prop_qmatrix})) and (b) is the corresponding intensity profiles at three $z$-locations. For a paraxial supergaussian beam ($w_0=$ \SI{56}{\um}, order = 50, $\lambda=$ \SI{400}{\nm}, nominal $M^2=4.0466$): (c) points are the beam widths $w_0$ calculated from the numerically propagated beam and the curve is the analytical width from Eq.~(\ref{eq:ISO}).}
    \label{fig:covariance}
  \end{center}
\end{figure}

For our first test beam, we use a linear combination of the first
four Hermite-Gauss (HG) modes with complex coefficients. The electric field distribution at the initial plane for the m-th HG mode is taken to be 
\begin{eqnarray}
    E_m\left(x\right) &=& n_m~H_m\left(\frac{\sqrt{2}x}{w_{0,m}}\right) e^{-\left(\frac{x}{w_{0,m}}\right)^2},  \\
    n_m &=& \left(\frac{2}{\pi}\right)^{1/4}\left(2^m m! w_{0,m}\right)^{-1/2},  \nonumber
\end{eqnarray}where $w_{0,m}$ is the beam waist of the mode and $H_m$ the m-th order Hermite polynomial. Although each HG mode is centered the total beam will generally be offset from the $x=0$ axis, so Eqs.~(\ref{eq:offset_Q},\ref{eq:abc_a}) in Appendix \hyperref[app:E2D_beams]{E} must be used. In addition, the $z$ location of the waist of a superposition will not generally be at the waist location of the individual HG modes. Accommodating these possibilities, we derive a completely general theoretical value of  $M^2$ for a superposition of HG modes with complex coefficients $c_n$ in Appendix \hyperref[app:GDerivation_M2_GeneralSuperposition]{F}, with Eq.~(\ref{eq:M2HG}) as the final result. 

The test beam's electric field is given by  $E\left(x\right) = \left(0.8+0.2i\right)E_0(x)+0.3E_1\left(x\right)+\left(-0.099-0.01i\right)E_2\left(x\right)+0.469E_3\left(x\right)$ with $w_{0,m}=$ \SI{642}{\um} for $m=0,1,2,3$. This particular superposition happens to have little wavefront curvature and, thus, Eq.~(85) for $M^2$ from \cite{weber1992propagation} is approximately correct. Both the latter equation and our general formula, Eq.~(\ref{eq:M2HG}), predict a value of $M^2=2.466$ for the above superposition.

For this first test beam we plot the elements of the offset $Q$ in Fig. \ref{fig:covariance}a. The element $a=w^{2}\left(z\right)/4$
(blue points) lies on the theoretical beam waist curve (Eq.~(\ref{eq:ISO}),
blue curve) and the other elements (yellow and black points) lie on
the curves given by the matrix elements of the propagated $Q$ matrix
(Eq.~(\ref{eq:prop_qmatrix}), yellow and black curves). This agreement
confirms that our numerical propagation of $E\left(x\right)$ is accurate
and our calculation of the covariance matrix $Q$ is correct. 

With the correctness of our numerical calculations established, we
now compare the ISO standard method to our covariance method for an offset beam to find $M^{2}$. The $M^2$ values found from our covariance method (Eqs.~(\ref{eq:M2_review},\ref{eq:abc_a})) at the
ten planes have average value $M^{2}=2.470$ with a standard deviation of $0.006$. Thus, as expected, our covariance method
results in the same value of $M^{2}$ and it is independent of the plane $z$
of the $E\left(x;z\right)$ used to calculate it. Moreover, it is in good agreement with the theoretical prediction of $M^{2}=2.466$. A fit to the waists in the ten planes, the ISO method, gives $M^{2}= 2.467$ with a fit error of $2.6 \times 10^{-10}$, again in good agreement. This suggests that using solely a single plane and the covariance method will agree with the ISO method up the second decimal place of $M^2$.

Our second test beam is meant to approximate the top hat intensity-distribution that would be transmitted by a slit.  In this way, it differs greatly from a basic Gaussian beam or even combinations of low-order HG modes.  Moreover, an exact top hat has an angular intensity distribution that follows a sinc-squared function and thus has a $\Delta \theta$ that diverges and is not well-defined, which makes it a particularly challenging test case. A one-dimensional supergaussian,  $E\left(x\right)=\exp[-\left(|x|/w_s\right)^N]$, of high order $N$ approximates a top hat function of full-width $2w_s$. Using  Eqs.~(\ref{eq:M2fromdet})  and (\ref{eq:abc}) we theoretically found $M^2=N\sqrt{\Gamma\left(3 / N\right)\Gamma\left(2-1/N\right)}/\Gamma\left(1 / N\right)\approx \sqrt{N/3}$, where the approximation is valid for large $N$ and $\Gamma$ is the usual gamma function. (Note, this formula differs from formulae in the literature, which are for the "cylindrical $M^2$", e.g., \cite{M2_supergaussian}.) The test beam is a supergaussian of order $N=50$ with equivalent slit width of $2w_s = \SI{0.1}{\mm}$  and a theoretical $M^2=4.0466$. Diffraction from such a slit has zero-to-zero angular width of $1.6 \times 10^{-2}$  $\SI{}{rad}$, putting it well within the paraxial regime. 

For this second test beam, we repeat the comparison of methods that was conducted on the first beam. The beam widths in the ten planes for the supergaussian are plotted in Fig. \ref{fig:covariance}c. The ten planes of our covariance method give an average value of $M^{2}$= 4.0462 with a standard deviation of $7.0 \times 10^{-5}$. The ISO method (based on a fit to the beam waists in multiple planes) gives $M^{2}= 4.0506$ with a fit error of $1.4 \times 10^{-9}$.  The ISO method's value only agrees with theory and covariance method up to two decimal places after rounding.  In contrast, our covariance method agrees with the analytical theoretical value to four decimal places, suggesting it is a more reliable method for this particularly challenging test case.

The success of our covariance method for finding $M^{2}$ with these two paraxial test beams validates its correctness and demonstrates that it stands in good agreement with the ISO protocol even for the extreme case of highly non-Gaussian beams. 

\section{Discussion and Conclusion \label{sec:four}}
This paper has focused on the simplest case, a paraxial coherent one-dimensional profile. In Appendix \hyperref[app:Dnon_paraxial]{D}, we similarly test non-paraxial beams. We show that both our covariance method and the ISO method fail for non-paraxial beams, as expected.  A one-dimensional description applies to simple coherent beams whose two-dimensional profile can be written as a product, $E\left(x\right)E\left(y\right)$, e.g., two dimensional HG modes. In Appendix \hyperref[app:E2D_beams]{E}, we generalize our covariance method to a wider variety of paraxial beams,  including beams with general two-dimensional profiles and incoherent beams.

In summary, we have demonstrated how one can compute the beam quality
factor $M^{2}$ as well as angular width, waist width, and waist location
for a beam from its electric field distribution in a single arbitrary
plane. The conventional method, prescribed in the ISO standard \cite{ISO-1},
relies on finding the beam width in ten prescribed propagation planes.
Since each width calculation requires three integrals of the intensity
distribution, the ISO method requires thirty integrals in total.
In comparison, our covariance method requires only six integrals of the electric
field distribution. Moreover, it eliminates the need to fit a function,
a key step in the ISO method. Further, our covariance method eliminates the need for an initial
estimate of the Rayleigh range and  waist location. 

Since there
are now many tools to measure the electric field profile of a beam,
our covariance method is amenable to use in laboratories and could be incorporated
in optical test and measurement products. That said, we envision our
method will be particularly useful for analyzing the electric field
output of optical simulations and could be directly built into common
simulation software. In this way, we expect our covariance method to calculate
beam quality will streamline optical research and development.

\section*{Acknowledgments}
We thank Orad Reshef for publicly posting a request for the method provided in this paper; the absence of solutions in the replies showed that the community was unaware of our covariance method.  This research was undertaken, in part, thanks to funding from the Natural Sciences and Engineering Research Council of Canada (NSERC), RGPIN-2020-05505, the Canada Research Chairs program, the Canada First Research Excellence Fund, and the MITACS Globalink Research Internship.

\section*{Supplemental material \label{SupplementaryMaterial}}
\href{https://figshare.com/s/a329aba5e892d1d29b6a}{Supplement material} provides the code used in this work.

\bibliography{M2_paper.bib}

\begin{thebibliography}{10}

\bibitem{siegman1990}
Anthony~E. Siegman.
\newblock {New developments in laser resonators}.
\newblock In Dale~A. Holmes, editor, {\em Optical Resonators}, volume 1224,
  pages 2 -- 14. International Society for Optics and Photonics, SPIE, 1990.

\bibitem{siegman1993defining}
Anthony~E Siegman.
\newblock Defining, measuring, and optimizing laser beam quality.
\newblock {\em Laser Resonators and Coherent Optics: Modeling, Technology, and
  Applications}, 1868:2--12, 1993.

\bibitem{Siegman1998}
Anthony~E Siegman.
\newblock How to (maybe) measure laser beam quality.
\newblock In {\em Diode Pumped Solid State Lasers: Applications and Issues},
  page MQ1. Optica Publishing Group, 1998.

\bibitem{ISO-1}
{Lasers and laser-related equipment - Test methods for laser beam widths,
  divergence angles and beam propagation ratios - Part 1}.
\newblock ISO 11146-1:2021(E).

\bibitem{digital_ImageFormation}
J.~W. Goodman and R.~W. Lawrence.
\newblock Digital image formation from electronically detected holograms.
\newblock {\em Applied Physics Letters}, 11(3):77--79, 1967.

\bibitem{Grilli:01}
S.~Grilli, P.~Ferraro, S.~De Nicola, A.~Finizio, G.~Pierattini, and R.~Meucci.
\newblock Whole optical wavefields reconstruction by digital holography.
\newblock {\em Opt. Express}, 9(6):294--302, Sep 2001.

\bibitem{boyd2019quantum}
R.W. Boyd, S.G. Lukishova, and V.N. Zadkov.
\newblock {\em Quantum Photonics: Pioneering Advances and Emerging
  Applications}.
\newblock Springer Series in Optical Sciences. Springer International
  Publishing, 2019.

\bibitem{hartmann1904objektivuntersuchungen}
Johannes Hartmann.
\newblock {\em Objektivuntersuchungen}.
\newblock Springer, 1904.

\bibitem{shack1971production}
Roland~V Shack.
\newblock Production and use of a lenticular {H}artmann screen.
\newblock In {\em Spring Meeting of Optical Society of America, 1971}, volume
  656, 1971.

\bibitem{schafer2002determination}
Bernd Sch{\"a}fer and Klaus Mann.
\newblock Determination of beam parameters and coherence properties of laser
  radiation by use of an extended {H}artmann-{S}hack wave-front sensor.
\newblock {\em Applied optics}, 41(15):2809--2817, 2002.

\bibitem{sharma2014wave}
Richa Sharma, J~Solomon Ivan, and CS~Narayanamurthy.
\newblock Wave propagation analysis using the variance matrix.
\newblock {\em JOSA A}, 31(10):2185--2191, 2014.

\bibitem{DigitalRecordingAndNumerical}
Thomas Kreis.
\newblock {\em Digital Recording and Numerical Reconstruction of Wave Fields},
  chapter~3, pages 81--183.
\newblock John Wiley \& Sons, Ltd, 2004.

\bibitem{schnars2014digital}
U.~Schnars, C.~Falldorf, J.~Watson, and W.~J{\"u}ptner.
\newblock {\em Digital Holography and Wavefront Sensing: Principles, Techniques
  and Applications}.
\newblock EBL-Schweitzer. Springer Berlin Heidelberg, 2014.

\bibitem{wielandy2007implications}
Stephan Wielandy.
\newblock Implications of higher-order mode content in large mode area fibers
  with good beam quality.
\newblock {\em Optics Express}, 15(23):15402--15409, 2007.

\bibitem{liao2009theoretical}
S~Liao, M~Gong, and H~Zhang.
\newblock Theoretical calculation of beam quality factor of large-mode-area
  fiber amplifiers.
\newblock {\em Laser physics}, 19:437--444, 2009.

\bibitem{Stutzki:14}
Fabian Stutzki, Florian Jansen, Hans-J\"{u}rgen Otto, Cesar Jauregui, Jens
  Limpert, and Andreas T\"{u}nnermann.
\newblock Designing advanced very-large-mode-area fibers for power scaling of
  fiber-laser systems.
\newblock {\em Optica}, 1(4):233--242, Oct 2014.

\bibitem{Zhou:15}
H.~Zhou, A.~Jin, Z.~Chen, B.~Zhang, X.~Zhou, S.~Chen, J.~Hou, and J.~Chen.
\newblock Combined supercontinuum source with >200 {W} power using a 3 $\times$
  1 broadband fiber power combiner.
\newblock {\em Opt. Lett.}, 40(16):3810--3813, Aug 2015.

\bibitem{hodgson1992beam}
N~Hodgson and T~Haase.
\newblock Beam parameters, mode structure and diffraction losses of slab lasers
  with unstable resonators.
\newblock {\em Optical and quantum electronics}, 24:S903--S926, 1992.

\bibitem{Jauregui:20}
Cesar Jauregui, Christoph Stihler, and Jens Limpert.
\newblock Transverse mode instability.
\newblock {\em Adv. Opt. Photon.}, 12(2):429--484, Jun 2020.

\bibitem{marshall2012handbook}
Gerald~F Marshall and Glenn~E Stutz.
\newblock {\em Handbook of optical and laser scanning}.
\newblock Taylor \& Francis, 2012.

\bibitem{forbes2014laser}
Andrew Forbes.
\newblock {\em Laser beam propagation: generation and propagation of customized
  light}.
\newblock CRC Press, 2014.

\bibitem{lavi1988generalized}
Shimon Lavi, Ron Prochaska, and Eliezer Keren.
\newblock Generalized beam parameters and transformation laws for partially
  coherent light.
\newblock {\em Applied optics}, 27(17):3696--3703, 1988.

\bibitem{weber1992some}
H~Weber.
\newblock Some historical and technical aspects of beam quality.
\newblock {\em Optical and Quantum Electronics}, 24:S861--S864, 1992.

\bibitem{du1992coherence}
K~M Du, G~Herziger, P~Loosen, and F~R{\"u}hl.
\newblock Coherence and intensity moments of laser light.
\newblock {\em Optical and quantum electronics}, 24:S1081--S1093, 1992.

\bibitem{siegman1991defining}
Anthony~E Siegman.
\newblock Defining the effective radius of curvature for a nonideal optical
  beam.
\newblock {\em IEEE Journal of Quantum Electronics}, 27(5):1146--1148, 1991.

\bibitem{gao2005characterization}
C~Gao, H~Weber, and M~Gao.
\newblock Characterization of laser beams by using intensity moments.
\newblock In {\em ICO20: Lasers and Laser Technologies}, volume 6028, pages
  428--435. SPIE, 2005.

\bibitem{Yoda:06}
Hidehiko Yoda, Pavel Polynkin, and Masud Mansuripur.
\newblock Beam quality factor of higher order modes in a step-index fiber.
\newblock {\em J. Lightwave Technol.}, 24(3):1350, Mar 2006.

\bibitem{Flamm:12}
Daniel Flamm, Christian Schulze, Robert Br\"{u}ning, Oliver~A. Schmidt, Thomas
  Kaiser, Siegmund Schr\"{o}ter, and Michael Duparr\'{e}.
\newblock Fast {M}$^2$ measurement for fiber beams based on modal analysis.
\newblock {\em Appl. Opt.}, 51(7):987--993, Mar 2012.

\bibitem{du2013real}
Yong-zhao Du, Guo-ying Feng, Hong-ru Li, Zhen Cai, Hong Zhao, and Shou-huan
  Zhou.
\newblock Real-time determination of beam propagation factor by
  {M}ach-{Z}ehnder point diffraction interferometer.
\newblock {\em Optics Communications}, 287:1--5, 2013.

\bibitem{du2016complex}
Yongzhao Du, Yuqing Fu, and Lixin Zheng.
\newblock Complex amplitude reconstruction for dynamic beam quality {M}$^2$
  factor measurement with self-referencing interferometer wavefront sensor.
\newblock {\em Applied Optics}, 55(36):10180--10186, 2016.

\bibitem{Schmidt:11}
Oliver~A. Schmidt, Christian Schulze, Daniel Flamm, Robert Br\"{u}ning, Thomas
  Kaiser, Siegmund Schr\"{o}ter, and Michael Duparr\'{e}.
\newblock Real-time determination of laser beam quality by modal decomposition.
\newblock {\em Opt. Express}, 19(7):6741--6748, Mar 2011.

\bibitem{Alonso2011}
M.~Alonso.
\newblock {W}igner functions in optics: describing beams as ray bundles and
  pulses as particle ensembles.
\newblock {\em Adv. Opt. Photon.}, 3:272--365, 2011.

\bibitem{crampin1986applicable}
M~Crampin and F.A.E. Pirani.
\newblock {\em Applicable Differential Geometry}.
\newblock Lecture note series. Cambridge University Press, 1986.

\bibitem{hazewinkel2012encyclopaedia}
M.~Hazewinkel.
\newblock {\em Encyclopaedia of Mathematics}.
\newblock Number v. 1 in Encyclopaedia of Mathematics. Springer Netherlands,
  2012.

\bibitem{dodonov1989universal}
VV~Dodonov and OV~Man'ko.
\newblock Universal invariants of paraxial optical beams.
\newblock {\em Computer Optics}, 1(1):65--68, 1989.

\bibitem{simon2000optical}
R~Simon and N~Mukunda.
\newblock Optical phase space, {W}igner representation, and invariant quality
  parameters.
\newblock {\em JOSA A}, 17(12):2440--2463, 2000.

\bibitem{dodonov2000universal}
Victor~V Dodonov and Olga~V Man’ko.
\newblock Universal invariants of quantum-mechanical and optical systems.
\newblock {\em JOSA A}, 17(12):2403--2410, 2000.

\bibitem{pettofrezzo1966matrices}
A.J. Pettofrezzo.
\newblock {\em Matrices and Transformations}.
\newblock Dover books on advanced mathematics. Prentice-Hall, 1966.

\bibitem{Bastiaans}
Martin~J. Bastiaans.
\newblock Second-order moments of the {W}igner distribution function in
  first-order optical systems.
\newblock In {\em Optical Society of America Annual Meeting}, page FC2. Optica
  Publishing Group, 1991.

\bibitem{ISO-3}
{Lasers and laser-related equipment - Test methods for laser beam widths,
  divergence angles and beam propagation ratios - Part 3 (ISO 11146-3: 2004)}.
\newblock ISO 11146-3:2004(E).

\bibitem{weber1992propagation}
H~Weber.
\newblock Propagation of higher-order intensity moments in quadratic-index
  media.
\newblock {\em Optical and Quantum Electronics}, 24:S1027--S1049, 1992.

\bibitem{M2_supergaussian}
Shirong Luo and Baida Lü.
\newblock M2 factor and kurtosis parameter of super-gaussian beams passing
  through an axicon.
\newblock {\em Optik}, 114(5):193--198, 2003.

\bibitem{f2f}
Charles~S. Adams and Ifan~G. Hughes.
\newblock {\em {Optics f2f: From Fourier to Fresnel}}.
\newblock Oxford University Press, 12 2018.

\bibitem{Astigmatic}
Andreas Letsch and Adolf Giesen.
\newblock {Characterization of a general astigmatic laser beam by measuring its
  ten second order moments}.
\newblock In Laurent Mazuray and Rolf Wartmann, editors, {\em Optical Design
  and Engineering II}, volume 5962. International Society for Optics and
  Photonics, SPIE, 2005.

\bibitem{ISO-2}
{ Lasers and laser-related equipment — Test methods for laser beam widths,
  divergence angles and beam propagation ratios — Part 2: General astigmatic
  beams (ISO 11146-2: 2021)}.
\newblock ISO 11146-2:2021(E).

\bibitem{floettmann2022coherent}
Klaus Floettmann.
\newblock Coherent superposition of orthogonal {H}ermite--{G}auss modes.
\newblock {\em Optics Communications}, 505:127537, 2022.

\end{thebibliography}
\bibliographystyle{unsrt}

\newpage

\section*{Appendix A: Paraxial propagation and beam parameters}\label{app:Aparaxial}
In this section we treat paraxial propagation within the matrix formalism
that we introduced in Section \ref{subsec:matrix_method}, which will allow us to determine
other key beam parameters and connect with the ISO standard procedure.
As discussed in Section \ref{subsec:matrix_method}, as the beam propagates in space, the
covariance matrix $Q$ evolves. The covariance matrix can be propagated
via the ray transfer matrix 
\begin{equation}
P=\begin{bmatrix}1 & z'\\
0 & 1
\end{bmatrix}\text{ ~~and the propagation law~~ }Q\left(z+z'\right)=P~Q\left(z\right)~P^{T},\label{eq:prop. matrix}
\end{equation}
where $T$ indicates transpose. This simple formula is valid in both
the Fresnel and Frauhnhofer diffraction regimes. The fact that $P$
has unit determinant ensures that the ellipse area is conserved under
paraxial propagation, which justifies Eq.~(\ref{eq:M2fromdet})
for all $z$-planes. 

We use Eq.~(\ref{eq:prop. matrix}) to propagate an arbitrary
$Q\left(z\right)$ by a distance $z'$,

\begin{equation}
Q\left(z+z'\right)\equiv Q'=\begin{bmatrix}a' & b'\\
b' & c'
\end{bmatrix}=\begin{bmatrix}a+bz'+\left(cz'+b\right)z' & b+cz'\\
b+cz' & c
\end{bmatrix}.\label{eq:prop_qmatrix}
\end{equation}
From this matrix we can retrieve the common beam parameters. First,
we note that $c$ is unchanged so, trivially, 
\begin{equation}
\Delta\theta=2\sqrt{c}.\label{eq:Angle_width_c}
\end{equation}
Next, we find the distance to the waist from $z$, the plane $Q$
was determined at, by setting $Q'$ to be diagonal, $b'=0=b+cz'$.
Taking the position of the waist ($Q'$) to be $z_{0}$ so that $z'=z-z_{0}\equiv-\Delta z$,
we then find
\begin{equation}
\Delta z=\frac{b}{c}=\frac{\expval{ x\theta+\theta x}}{2\expval{\theta^{2}}},\label{eq:waist_location}
\end{equation}
where a positive $\Delta z$ means  the waist is further along
the beam propagation direction from the plane that $E\left(x;z\right)$ (and
$Q$) was determined in. With $z'=-b/c$ substituted in $Q'$ Eq.~(\ref{eq:prop_qmatrix}), we find the beam waist,
\begin{equation}
w_{0}=2\sqrt{a'}=2\sqrt{a-b^{2}/c}.\label{eq:waist_Q}
\end{equation}

We now instead set $Q\left(z_{0}\right)$ to be the waist and, thus, diagonal
($b=0$), and using $M^{2}$ from Eq.~(\ref{eq:M2basic}), the $a'$
element of $Q'$ gives the beam width: 
\begin{equation}
4a'=w\left(z\right)=w_{0}^{2}+\left[\frac{M^{2}\lambda}{\pi w_{0}}\right]^{2}\left(z-z_{0}\right)^{2},\label{eq:ISO}
\end{equation}
where $z'=z-z_{0}$. This gives the Rayleigh range, 
\begin{equation}
z_{R}=\frac{\pi w_{0}^{2}}{M^{2}\lambda}=\frac{\lambda}{4\pi c}.\label{eq:rayleigh_range_c}
\end{equation} In other words, the propagation of an arbitrary
paraxial beam is same as the one of a Gaussian beam magnified by $M^{2}$.

Equation (\ref{eq:ISO}) above is precisely the relation that the ISO standard uses to model
the evolution of the beam width under spatial propagation. One first
finds the intensity distributions in ten planes, then
finds their widths, and then fits them to the hyperbolic function
in Eq.~(\ref{eq:ISO}). This fit yields values for $M^{2}$, the beam waist $w_{0}$ and its location $z_{0}$ \cite{ISO-1}. A big disadvantage of this approach is that we already
need to have rough values for the position of the beam waist as well
as the $M^{2}$ parameter in order to compute the effective Rayleigh
distance $z_{R}$ and be able to follow the steps given in the protocol.

\section*{ Appendix B: Propagation simulation}\label{app:BFT_prop}

For the spatial propagation of the two beams to the ten planes, we draw on the angular spectrum method \cite{f2f}. The Fourier transform
of $E\left(x;0\right)$ decomposes the input field into plane waves, which then
can be propagated by distance $z$ via: 
\begin{equation}
E\left(x;z\right)=\mathcal{F}^{-1}\left[e^{ik_{z}z}\mathcal{F}\left[E\left(x;0\right)\right]\right],\label{eq:angular_spectrum}
\end{equation}
where $k_{z}=\sqrt{k^{2}-k_{x}^{2}}$. This method is accurate for
all angles for a scalar field. That is, it is valid beyond the paraxial
approximation. For the implementation, we used the native Fast Fourier
Transform (FFT) in Python. 

\section*{ Appendix C: Numerical details  \label{app:CNumericalDetails} }

We now briefly discuss some details of the numerical implementation
for our covariance method. We use Python to evaluate Eq.~(\ref{eq:abc}).
Derivatives were calculated with the Numpy gradient function, which
uses the central difference method. Scipy integrate (Simpson's rule)
was used for integrals. Code is provided in the \hyperref[SupplementaryMaterial]{ supplementary material}.

The accuracy of our covariance method is mainly set by the $x$-position grid
of the electric field profile $E\left(x\right)$. This directly sets the range and grid density in $x$ and indirectly sets them in $\theta$.  In particular, since the two directions are related by a Fourier transform, the grid spacing $\delta \theta $ in the $\theta$-direction is set by the range $L$ in the $x$-direction, $\delta \theta \propto 1/\left(kL\right)$. Additionally, the range $\Theta$ in the $\theta$-direction is set by the grid spacing $\delta x$ in the $x$-direction, $\Theta\propto 1/\left(k\delta x\right)$. 

The two goals are to have sufficient
grid density to resolve the state and sufficient range to span the
state in phase-space. That is, in both the $x$ and $\theta$ directions, we aim for $r$ points across the minimum width of the state and a range of $s$ times the maximum width of the state.  The minimum and maximum widths in $x$ are $w_0$ and $w\left(z\right)$, which set  $\delta x<w_{0}/r$ and $L>w\left(z\right)s$. The beam width is always $\Delta \theta$  in $\theta$, so $\delta \theta = \Delta \theta/r$ and $\Theta=s\Delta \theta $. By the Fourier relations from above, these set $L>\Delta\theta/\left(rk\right)$ and $\delta x<\left(s\Delta\theta\right)/k$, respectively. These four inequalities ensure sufficient resolution and range.

The inequalities determine the grid for a given beam state based on the chosen range and resolution parameters. For the latter, we suggest, $s=r=10$. Since $w_{0}$ and $\Delta\theta$ are not
known a priori, one should start with a trial grid $L>10$ $ w\left(z\right)$ and
then calculate $w_{0}$ and $\Delta\theta$ from Eqs.~(\ref{eq:abc},\ref{eq:parameters}). At this point, one would potentially need to revise the grid. Alternately, if $E\left(x\right)$ is from an optical simulation and computational power is not a constraint, setting $\delta x<\lambda/10$ will certainly be sufficient since it is five times better than the Nyquist limit. Satisfying the above conditions will ensure an accurate value of $M^{2}$.

\section*{Appendix D: Non-paraxial beams \label{app:Dnon_paraxial} } 

Both the ISO protocol and our covariance method rely on the paraxial
approximation. More problematically, the derivation of the $M^{2}$ from
the uncertainty principle relies on the small-angle
approximation, so it is not clear that the definition is completely general.
In this section, we briefly examine the validity of both the ISO and covariance methods in
the non-paraxial regime. For the following, $\lambda= \SI{400}{\nm}$ and we
use position range of length $L= \SI{200}{\mm}$   and a position grid density
$\delta x=\lambda/2$.

\begin{figure}[hbt!]
  \begin{center}
  Numerical and analytic propagation of a non-paraxial supergaussian beam \\
\hspace{2mm}  (a) \hspace{45mm} (b)\\
  \includegraphics[width=6cm, height=4.0cm]{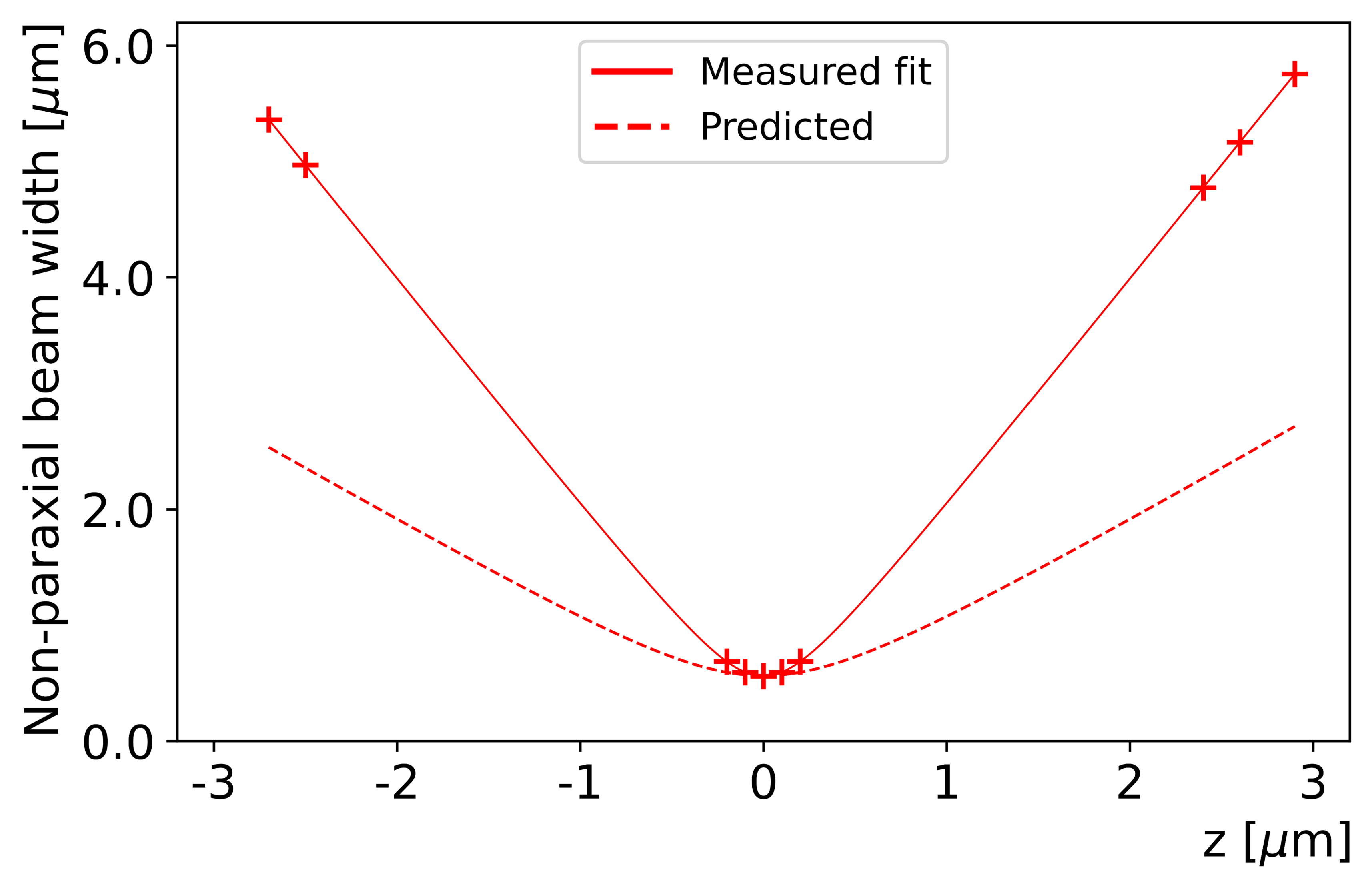}
  \includegraphics[width=6 cm, height=4.0cm]{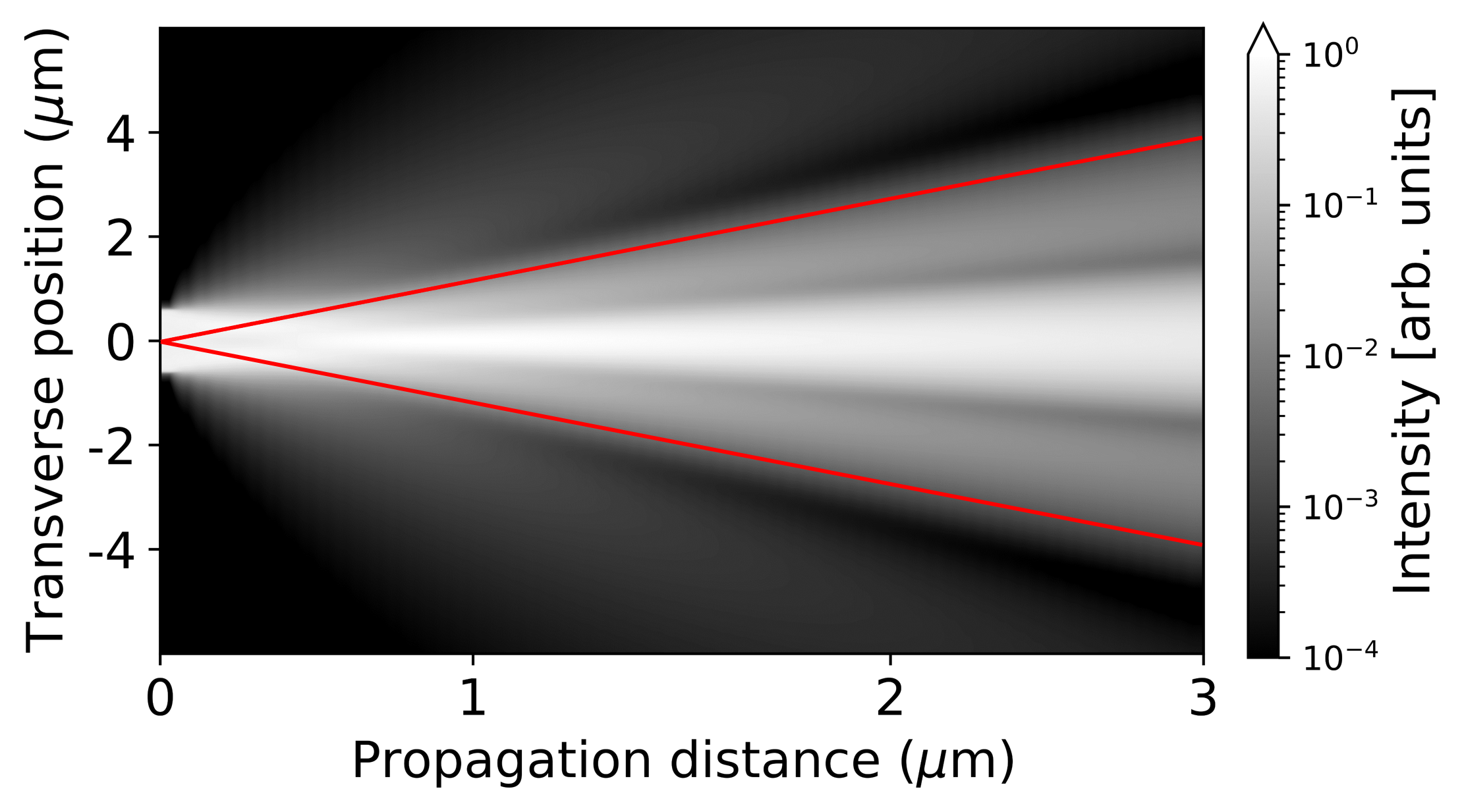}\\
  \caption{(a) Width evolution of a non-paraxial supergaussian ($w_0=$ \SI{0.56}{\um}, order = 50, $\lambda=$ \SI{400}{\nm}). Measured widths (indicated by points) do not lie on predicted curve for $M^2=4.05$. (b) Intensity plot of the same supergausssian. Red lines indicate the full angular spread of the beam $2\Delta \theta =  \SI{0.91}{rad}$.}
    \label{fig:non_paraxial}
  \end{center}
\end{figure}

To examine the non-paraxial regime, we decrease the effective slit
full-width of the order $N=50$ supergaussian to 0.5 $\mu$m, which increases
the zero-to-zero angular spread to $\SI{1.6}{rad}$ putting it outside the paraxial approximation. Fig. \ref{fig:non_paraxial}b should
give an impression of the non-paraxiality of the supergaussian beam.
The beam quality factor for a supergaussian is solely a function of order
so theoretical beam quality should be the same as the $N=50$ supergaussian from  Section \ref{subsec:comparison}, $M^{2} = 4.0466$. Applying the ISO method, the width measurements
can still be fitted to a hyperbolic function (Fig. \ref{fig:non_paraxial}a,
top curve). However, the resulting $M^{2}$ factor more than doubles to
a value of 9.28. While the covariance matrix gives the correct result for $M^2$ in the $z=0$-plane, the values quickly diverge as soon as propagation starts. At $z=$ \SI{2.9}{\um}, the covariance method gives, $M^2=43.5$, which is off by a factor of five at least. The limitation to paraxial beams might be lifted when defining
the beam matrix through a non-paraxial version of the Wigner function
\cite{Alonso2011}, but a thorough investigation would be in order
here.

\section*{Appendix E: General two-dimensional beams \label{app:E2D_beams} }
We now discuss some potential generalizations of our covariance method
to a wider class of paraxial optical beams. We leave the details for future work.

\subsection*{Offset beams}
In some optical simulations and all laboratory measurements one cannot assume that beam is perfectly centered on the beam axis, i.e., $\expval{x}\neq0$ and $\expval{\theta}\neq0$. In this case,  the covariance matrix takes the more general form:
\begin{equation}
Q\left(z\right)=\begin{bmatrix} \expval{x^2} - \expval{x}^{2} & \frac{1}{2} \expval{x\theta+\theta x } -\expval{ x } \expval{ \theta}\\
\frac{1}{2}\expval{ x\theta+\theta x} -\expval{ x } \expval{ \theta}  & \expval{ \theta^2 }-\expval{ \theta}^{2} 
\end{bmatrix}=\begin{bmatrix}a & b\\
b & c
\end{bmatrix}.\label{eq:offset_Q}
\end{equation}
The corresponding generalized formulas for the matrix elements are
\begin{eqnarray}
a & = & \frac{1}{n}\int{x^{2}|E|^{2}dx}-\left(\frac{1}{n}\int{x|E|^{2}dx}\right)^{2},\label{eq:abc_a}\\
b & = & \frac{i\lambda}{2\pi}\Biggl[\left(\frac{1}{n}\int{x|E|^{2}dx}\right)\left(\frac{1}{n}\int{E^{*}\pdv{E}{x}dx}\right)-\left(\frac{1}{2}+\frac{1}{n}\int{xE^{*}\pdv{E}{x}dx}\right)\Biggr],\nonumber \\
c & = & \left[\frac{\lambda}{2\pi}\right]^{2}\Biggl[\left(\frac{1}{n}\int{E^{*}\pdv{E}{x}dx}\right)^{2}-\frac{1}{n}\int{E^{*}\pdv[2]{E}{x}dx}\Biggr],\nonumber \\
n & = & \int{|E|^{2}dx}.\nonumber 
\end{eqnarray}
As before, these can be used to find $M^2$ and other beam parameters using Eqs.~(\ref{eq:M2_review},\ref{eq:parameters}).

\subsection*{Simple astigmatic beams}
So far, we have presented our
method in one transverse dimension $x$ for clarity. This one-dimensional
analysis is straightforward to generalize to two-dimensional simple astigmatic beams; those
that have have circular and elliptical cross sections, respectively. To apply our covariance method to simple astigmatic beams, one first would find
the principal axes of the elliptical transverse profile and denote these by $x$ and $y$. Along these directions, the total field is separable, $E\left(x,y;z\right)=E_{x}\left(x;z\right)E_{y}\left(y;z\right)$,
and each transverse direction can be separately analyzed by our covariance
method. The result would be two values, $M_{x}^{2}$ and $M_{y}^{2}$,
from which one can define an effective beam quality $M_{\text{eff}}^{2}=\sqrt{M_{x}^{2}M_{y}^{2}}$. See \cite{ISO-1} for details.

\subsection*{General astigmatic beams}
We now propose a route to adapt our covariance method to find the
beam quality of more general two-dimensional paraxial fields, $E\left(x,y;z\right)\neq E_{x}\left(x;z\right)E_{y}\left(y;z\right)$.
Unlike simple astigmatic beams, the principal axes of their transverse
cross sections can rotate while the beam is propagating \cite{Astigmatic}.
Part 2 of the ISO standard \cite{ISO-2} derives an effective beam quality
$M_{\text{eff}}^{2}$ in terms of moments of the Wigner function in
the four-dimensional phase-space. Essentially, the state is a generalized
ellipse in this four-dimensional space and $M_{\text{eff}}^{2}$ is
the area of that state. Analogous to the generalized $P$ matrix in Part 2 of the ISO standard, a generalized $Q_{xy}\left(z\right)$ matrix would be a 4x4 matrix:
\begin{equation}
\ Q_{xy}\left(z\right)\equiv\left[\begin{array}{cc}
Q_{x} & D_{xy}\\
D_{xy} & Q_{y}
\end{array}\right].\label{eq:gen_Q}
\end{equation}
Here, the block diagonals are the standard 2x2
$Q$ matrices Eq.~(\ref{eq:covariance matrix}):  $Q_{x}\equiv Q$ and $Q_{y}$ is analogous.
The $x$ and $y$ are arbitrary orthogonal
transverse directions unconnected to the beam state. The two off-diagonal blocks are identical and
given by
\begin{equation}
D_{xy}\left(z\right)=\left[\begin{array}{cc}
\expval{ xy } & \expval{ x\theta_{y} }\\
\expval{ y\theta_{x}} & \expval{ \theta_{x}\theta_{y} }
\end{array}\right]. \label{eq:D_matrix}
\end{equation}
Note, the $D$ blocks involve one variable from each direction and thus
do not require the two orderings (e.g., $\expval{ x\theta+\theta x}$)
that are essential for the 2x2 $Q$ matrix. With this generalized
covariance matrix, $M_{\text{eff}}^{2}=\frac{4\pi}{\lambda}\sqrt[4]{\det Q_{xy}\left(z\right)}$.
Unlike the $P$  matrix in the ISO procedure, $Q_{xy}\left(z\right)$ is found directly from
the electric field $E\left(x,y;z\right)$.

\subsection*{Incoherent beams}
We now consider generalizing our covariance method to incoherent beams.
We tested our covariance method with beams that are coherent superpositions of
HG modes. If the beam is an incoherent mixture of beams, one would
need to use a density matrix formalism, or equivalently, a coherence
matrix to represent the beam in single $z$ plane, $\rho\left(x,x'\right)\equiv\overline{E\left(x\right)E^{*}\left(x'\right)}$.
Here, the bar indicates an ensemble or time average, as is standard
in statistical optics. We leave the details for future work, but in short the
expectation values in $Q_{xy}\left(z\right)$ would then be $\expval{ v}=\mathrm{Tr}\left[v\rho\right]$
and the beam quality $M^{2}$ would be found from $Q$ as before,
from Eq.~(\ref{eq:M2fromdet}). This approach would complete the presented covariance method. However previous work have shown how to obtain $M^2$ for an incoherent mixture of beams \cite{siegman1990,du1992coherence}, we refer the reader to such references for further details. 

\section*{Appendix F: Derivation of $M^2$ for a general superposition of HG beams \label{app:GDerivation_M2_GeneralSuperposition} }
We now derive a general formula for $M^2$ for a beam $E\left(x\right)$ that is a superposition of HG beams $E_m\left(x\right)$. This result is more general than the equations obtained in Refs. \cite{weber1992some, du1992coherence,floettmann2022coherent}. Mathematically $E_m\left(x\right)$ also describes the wave function of the energy states of a simple harmonic oscillator in quantum mechanics. Thus we use the \textit{bra-ket} mathematical formalism to calculate expectation values as is standard in quantum mechanics. Consider a general state $\ket{\psi}$ described in the HG beam basis i.e., $\ket{\psi} = \sum_{n=0}^{\infty} c_n\ket{n}$ where $\ket{n}$ is the state of the HG mode of order n and $c_n$ are complex coefficients satisfying  $ \sum_{n=0}^{\infty} |c_n|^2 = 1.$

In the main text we found Eq.~(\ref{eq:M2fromdet}) as the basis of the covariance method. We use the general form for the covariance matrix elements in Appendix \hyperref[app:E2D_beams]{E}, Eq.~(\ref{eq:abc_a}) to find $M^2$ from $\ket{\psi}$. It is useful to use dimensionless position $X$ and momentum $P$ (rather than angle) operators defined as follows $X = x/2\sigma_x$ and $P = p\sigma_x/\hbar = 2\pi\sigma_x\theta/\lambda.$ In terms of these dimensionless operators, Eq.~(\ref{eq:M2fromdet}) is rewritten as 
\begin{eqnarray}
    M^2 & = & 4\sqrt{\det Q} \label{eq:M2_DimensionlessXP} \\ 
     & = & 4\sqrt{ \left(\Delta X\right)^2\left(\Delta P\right)^2 - \left( \expval{XP+PX }/2 - \expval{ X } \expval{P} \right)^2  }  \nonumber \\
     & = & 4\sqrt{ \left(\expval{ X^2 }-\expval{ X}^{2}\right)\left(\expval{ P^2 }-\expval{ P}^{2}\right) - \left( \expval{XP+PX }/2 - \expval{ X } \expval{P} \right)^2 }  \nonumber \\
     & = & 4\sqrt{\expval{ X^2 }\expval{ P^2 } - \expval{ X^2 }\expval{ P }^2 -  \expval{ X }^2\expval{ P^2}  + 2 \expval{ X }\expval{ P }\Re \expval{ XP } - \left(\Re\expval{ XP}\right)^2}, \nonumber
\end{eqnarray}
where we have used $\frac{1}{2}\expval{XP+PX} = \Re\expval{ XP }.$

The calculation of the expectation values appearing in above equation can be done using the ladder operators $a$ and $a^\dagger$ operators commonly used in quantum mechanics. These operators are related to $X$ and $P$ according to the following equations $X = \left( a + a^\dagger \right)/2$ and  $P = i\left( a^\dagger - a \right)/2$. The action of these operators on the $\ket{n}$ states is given by $a\ket{n} = \sqrt{n}\ket{n-1}$ and $a^\dagger\ket{n} = \sqrt{n+1}\ket{n+1}$. Expectation values can now be easily obtained, for example $\expval{P}$ is calculated as follows:

\begin{eqnarray}
    \expval{ P } & = &  \expval{  i\left( a^\dagger - a \right)/2 } \nonumber \\
& = & \frac{i}{2}\sum_{n=0}^{\infty} \left( c_nc_{n+1}^* - c_n^*c_{n+1} \right) \sqrt{n+1} \nonumber \\
& = & -\Im\left(B\right),
\end{eqnarray} 
where constants $A$,$B$, $C$ will be defined explicitly at the end. While $\expval{P^2}$ is given by 

\begin{eqnarray}
    \expval{P^2} & = & -\expval{a^2 + \left(a^\dagger\right)^2 - 2\expval{n} - 1}/4 \nonumber \\
 & = & -\frac{1}{2} \sum_{n=0}^{\infty} \Re\left(c_nc^*_{n+2}\right)\sqrt{ \left(n+1\right)\left(n+2\right)} + 1/2\sum_{n=0}^{\infty}n|c_n|^2 + 1/4 \nonumber \\
 & = & -\frac{1}{2}\Re\left(A\right) + \frac{C}{4}.
\end{eqnarray}
Following a similar procedure, one obtains  $\expval{X} = \Re\left(B\right)$, $\expval{X^2} = \frac{1}{2}\Re\left(A\right) + \frac{C}{4} $, and $\Re\left(XP\right)=-\frac{1}{2}\Im\left(A\right)$.

Substituting these into Eq.~(\ref{eq:M2_DimensionlessXP}) we obtain the following general expression of $M^2$ for an arbitrary superposition of HG beams:

\begin{equation}
M^2  =   \sqrt{C^2  + 8\Re\left(B^2A^* \right) - 4|A|^2 - 4|B|^2C},
\label{eq:M2HG}
\end{equation}
where $A^*$ is the complex conjugate of $A$ and 
\begin{eqnarray}
    A & = & \sum_{n=0}^\infty c_nc^*_{n+2}\sqrt{ \left(n+1\right)\left(n+2\right)},\\
    B & = & \sum_{n=0}^\infty  c_nc^*_{n+1} \sqrt{n+1},\\
    C & = & 1 + 2\sum_{n=0}^\infty n|c_n|^2.
\end{eqnarray} 

\end{document}